\documentclass[11pt,a4paper]{scrartcl}
\pdfoutput=1

\usepackage[utf8]{inputenc}
\usepackage[T1]{fontenc}
\usepackage{lmodern}

\usepackage{graphicx}
\usepackage{microtype}

\usepackage{float}
\usepackage{cite}
\usepackage{amssymb}
\usepackage{amsmath}
\usepackage{url}
\usepackage[dvipsnames]{xcolor}
\usepackage{cancel}
\usepackage{multirow}
\usepackage[normalem]{ulem}
\usepackage[toc,page]{appendix}
\usepackage[onehalfspacing]{setspace}
\usepackage{booktabs}
\usepackage[format=plain,labelfont={bf},font={small}]{caption}
\usepackage[colorlinks,citecolor=blue,urlcolor=blue,linkcolor=blue]{hyperref}

\setkomafont{disposition}{\normalcolor\normalfont\bfseries}

\newcommand{\MSbar}{\overline{\text{MS}}}
\newcommand{\alphas}{\alpha_{\text{s}}}

\newcommand{\refeq}[1]{Eq.~\eqref{#1}}
\newcommand{\refeqs}[1]{Eqs.~\eqref{#1}}

\newcommand{\reffig}[1]{Fig.~\ref{#1}}

\newcommand{\Reffig}[1]{Figure~\ref{#1}}

\newcommand{\totd}{{\mathrm{d}}}

\newcommand{\mr}{\mathrm}
\newcommand{\mc}{\mathcal}
\newcommand{\mur}{\mu_{\text{R}}}

\newcommand{\Mmed}{M_{\text{med}}}
\newcommand{\MV}{M_V}
\newcommand{\MQ}{M_{\tilde Q}}
\newcommand{\mchi}{m_{\chi}}

\newcommand{\beq}{\begin{equation}}
\newcommand{\eeq}{\end{equation}}
\newcommand{\bea}{\begin{eqnarray}}
\newcommand{\eea}{\end{eqnarray}}

\newcommand{\tree}{{\text{tree}}}
\newcommand{\oneloop}{{\text{1-loop}}}
\newcommand{\CT}{{\text{CT}}}
\newcommand{\nlo}{{\text{NLO}}}
\newcommand{\EFT}{{\text{EFT}}}

\newcommand{\si}{{\text{SI}}}
\newcommand{\sd}{{\text{SD}}}

\newcommand{\sisi}{\sigma_p^\si}
\newcommand{\sisd}{\sigma_n^\sd}

\newcommand{\micromegas}{\textsc{micrOMEGAs}}

\begin{document}

\renewcommand*{\thefootnote}{\fnsymbol{footnote}}

\begin{center}
	{\Large \textbf{Direct detection of dark matter:\\[1ex]
	Precision predictions in a simplified model framework}}\\
	\vspace{.7cm}
	Christoph Borschensky\footnote{\texttt{christoph.borschensky@uni-tuebingen.de}},
	Gabriele Coniglio\footnote{\texttt{gabriele.coniglio@uni-tuebingen.de}},
	Barbara J\"ager\footnote{\texttt{jaeger@itp.uni-tuebingen.de}},
	Josef Jochum\footnote{\texttt{josef.jochum@uni-tuebingen.de}},
	Vincent Schipperges\footnote{\texttt{vincent.schipperges@uni-tuebingen.de}},

	\vspace{.3cm}
	\textit{
		University of T\"ubingen, Auf der Morgenstelle 14, 72076 T\"ubingen, Germany\\
	}
\end{center}

\renewcommand*{\thefootnote}{\arabic{footnote}}
\setcounter{footnote}{0}

\vspace*{0.1cm}
\begin{abstract}
	We present a calculation of the next-to-leading order QCD corrections for the scattering of Dark Matter particles off nucleons in the framework of simplified models with $s$- and $t$-channel mediators. These results are matched to the Wilson coefficients and operators of an effective field theory that is generally used for the presentation of experimental results on spin-independent and spin-dependent direct detection rates.
	Detailed phenomenological studies illustrate the complementary reach of collider searches for Dark Matter and the direct detection experiments CRESST and XENON. In the case of cancellation effects in the tree-level contributions, one-loop corrections can have a particularly large impact on exclusion limits in the case of combined $s + t$-channel models.
\end{abstract}

\tableofcontents

%%%%%%%%%%%%%%%%%%%%%%%%%%%%%%%%%%%%%%%%%%%
\section{Introduction}\label{s:intro}
%%%%%%%%%%%%%%%%%%%%%%%%%%%%%%%%%%%%%%%%%%%
The field of elementary particle physics has entered a new era in 2012: With the discovery of the Higgs boson at the CERN Large Hadron Collider (LHC)~\cite{Aad:2012tfa,Chatrchyan:2012ufa} a long-sought entity of nature has been established.  Yet, a number of puzzles remain that are not accounted for by the Standard Model (SM) of elementary particles. Among these the quest for the origin of Dark Matter (DM) takes a prominent role.
Astrophysical observations \cite{Bertone:2004pz} point towards the existence of non-baryonic matter which is subject to the gravitational force. However, apart from its gravitational properties very little is known about the nature of DM.
In the context of particle physics, it is often assumed that DM consists of a new type of particle, not accounted for within the SM~\cite{Bertone:2004pz}. Prominent examples of such postulated particles are new types of neutrinos \cite{Ma:2006km,Klasen:2013jpa}, axions -- light particles introduced to resolve the strong CP problem of QCD \cite{Peccei:1977hh,Peccei:1977ur} that could account for DM if their masses were in the meV range~\cite{Ipser:1983mw} -- or weakly interacting massive particles (WIMPs) with a mass in the few-hundred GeV range (see, e.g., Ref.~\cite{Arcadi:2017kky} for a recent review).
Such WIMPs emerge, for instance, in supersymmetric models where the lightest stable supersymmetric particle is constituted by a neutralino. More recently there have also been attempts to account for WIMPs in a more generic way by the construction of so-called {\em simplified models} which aim at capturing the main features of these new particles and their interactions, in particular their mass, spin, and couplings to a specific mediator that in turn couples to SM particles (see, e.g., Ref.~\cite{Abdallah:2015ter} for a recent discussion of several simplified models in the context of LHC searches). While not providing an ultimately UV-complete theory, such models have the advantage of being simple to use and featuring a relatively small number of parameters that are to be determined from experimental measurements. Once data have constrained the parameters of such a simplified model, theorists can use this input for the construction of more sophisticated extensions of the SM.

From the experimental side, this approach requires ways to constrain mass, spin, and coupling strengths of the DM candidate and the mediator particle. Ideally, such information is extracted from and cross-checked among conceptually entirely different types of searches. Indeed, one can distinguish three major types of experiments searching for Dark Matter:
Indirect, astrophysical searches aiming to detect SM signatures resulting from DM annihilation processes; searches for DM production at high-energy colliders; and direct detection experiments that are designed to identify the recoil a DM particle causes in a nuclear target. Here, we will concentrate on the latter.
Depending on the design of a specific direct detection experiment, its sensitivity is typically tailored to a particular mass range of the DM candidate: Detectors making use of dual-phase time projection chambers are particularly effective in the search for DM particles with a mass above about 10~GeV~\cite{Undagoitia:2015gya}.
Cryogenic experiments are most sensitive to very light DM candidates, with a mass below a few GeV, because of their low energy threshold~\cite{Undagoitia:2015gya}.

The recoil rate in direct detection experiments depends on several quantities of astrophysical origin, such as the local DM density and the velocity distribution of DM, as well as nuclear properties of the detector material and the nuclear scattering cross section. For a multitude of simplified models, the leading expressions for the scattering cross sections can be obtained from the literature (see, e.g., Ref.~\cite{Dent:2015zpa}). In order to improve the theoretical predictions for the recoil rate, the uncertainties on all ingredients must be reduced. In particular, the calculation of the scattering cross section can be improved with conventional methods in particle physics, i.e.\ by including higher-order corrections in the perturbative expansion. For specific models, such higher-order terms as well as loop-induced contributions have been discussed, for instance in \cite{Drees:1992rr,Drees:1993bu,Hisano:2010ct,Haisch:2013uaa,Crivellin:2014qxa,Abe:2015rja,Hisano:2015bma,Hisano:2015rsa,Klasen:2016qyz,Abe:2018emu,Azevedo:2018exj,Ishiwata:2018sdi,Ghorbani:2018pjh,Ertas:2019dew,Mohan:2019zrk,Glaus:2019itb,Glaus:2020ape}.

In this work, we aim to provide accurate predictions for direct detection rates in the context of simplified models, including next-to-leading order (NLO) QCD corrections to the genuine tree-level processes where a DM particle scatters off a (anti-)quark from the nucleon.
Including these perturbative corrections will allow us to obtain results with theoretical uncertainties related to unknown higher-order corrections reduced as compared to pure tree-level estimates. In order to assess remaining perturbative uncertainties in the considered observables we perform a detailed assessment of the dependence on unphysical scales.
We show representative predictions within specific simplified models and discuss how exclusion limits of direct detection experiments can be used to derive constraints on model parameters such as the mass of the DM particle and mediator, or coupling strengths. We furthermore  investigate the complementarity of results from direct detection experiments to those obtained at colliders, in particular recent measurements by the CRESST~\cite{Abdelhameed:2019hmk} and XENON~\cite{Aprile:2018dbl,Aprile:2019dbj} experiments, and at the LHC~\cite{Aaboud:2017phn}.

In Sec.~\ref{s:theory} we describe the simplified model framework we are using as the basis of this study. We then explain how to connect these underlying models based on the properties of elementary particles to experimentally accessible direct detection rates in a nuclear medium. Numerical results are presented in Sec.~\ref{s:results}, followed by our conclusions in Sec.~\ref{s:conclusion}. Technical details of our calculation are summarized in \ref{s:NLO-expressions}.

%%%%%%%%%%%%%%%%%%%%%%%%%%%%%%%%%%%%%%%%%%%
\section{Theory overview}\label{s:theory}
%%%%%%%%%%%%%%%%%%%%%%%%%%%%%%%%%%%%%%%%%%%
In the following, we introduce the DM models that we are using in our analyses and describe the steps to obtain predictions for DM-nucleon cross sections, as relevant for direct detection experiments, up to NLO-QCD accuracy.

%%%%%%%%%%
\subsection{Simplified dark matter models}
\label{s:theory:models}
A particularly promising candidate for a DM particle is constituted by a -- as of yet hypothetical -- weakly interacting massive particle (WIMP) with a mass in the few-hundred GeV range that is subject only to the gravitational and weak interactions. Such a WIMP occurs in many extensions of the Standard Model, for instance models with additional Higgs doublets or in supersymmetric theories where the lightest neutralino provides a natural DM candidate. 

Explicit calculations of experimentally accessible quantities (such as production rates and cross sections) can be performed within any such model. However, it can be advantageous to provide predictions of a more general nature that are not relying on the many parameters of a complex theory, such as supersymmetry, but only depend on those features that govern the interaction of the DM particle with the SM in the considered environment. To this end, two different strategies have been devised: The framework of effective field theories (EFTs), or so-called {\em simplified models}.
Assuming that the interaction of the DM candidate with SM particles proceeds exclusively via a specific mediator particle, in the kinematic range where the mass of the mediator is much higher than all other energy scales of the considered reaction, in the EFT approach the heavy mediator particle is simply integrated out. Interactions between the SM and the DM sector can then be expressed in terms of effective operators.
More control on details of a specific reaction is retained in simplified models. Such models are designed to capture the main features of reactions that are sensitive to the DM particle itself and the mediator between the SM and the DM sector. Contrary to the EFT approach, the dependence on the mediator particle is fully retained. Basic simplified models are constructed under the assumption that there is only a single type of DM particle, and a single type of mediator.

In this work, we consider simplified models with a single fermionic DM candidate. The interaction between the SM and the DM fields proceeds either via the exchange of a neutral massive vector or axial-vector mediator, or via the exchange of a color-charged scalar mediator. Because of the characteristic topologies of the dominant DM production modes associated with these models at colliders, the models with neutral or charged mediators are often referred to as {\em ``$s$-channel'' or ``$t$-channel'' simplified models}, respectively.

We use the same notation for the terms in the Lagrangian as in our previous work \cite{Borschensky:2018zmq}. For the simplified $s$-channel model we consider the interaction terms in the Lagrangian are of the form \cite{Dudas2009}
\begin{align}
\label{eq:L-sim-mod-s}
	\mathcal{L}_{V} &= \bar{\chi}\gamma^\mu\left[g_\chi^V-g_\chi^A\gamma_5\right]\chi V_\mu + \sum_q \bar{q}\gamma^\mu\left[g_q^V-g_q^A\gamma_5\right]q V_\mu\,,
\end{align}
where $V$ denotes the vector mediator field with mass $\MV$, $\chi$ the fermionic DM field with mass $\mchi$, $q$ a quark field with mass $m_q$ for each quark flavor $q = u,d,s,c,b,t$, and $g_f^{V(A)}$ the coupling strength between the vector (axial-vector) mediator and fermionic particle~$f$ (with $f = q,\chi$).

For the $t$-channel models we assume a Lagrangian of the form \cite{An:2013xka,DiFranzo:2013vra,Abercrombie:2015wmb,Ko:2016zxg}
\begin{align}
\label{eq:L-sim-mod-t}
	\mathcal{L}_{\tilde{Q}} &= -\Big[\lambda_{Q_L} \bar{\chi}\left(\tilde{Q}^\dag_L \cdot Q_L\right) + \lambda_{u_R}\tilde{Q}^*_{u_R}\bar{\chi} u_R + \lambda_{d_R}\tilde{Q}^*_{d_R}\bar{\chi} d_R + \mathrm{h.c.}\Big]\notag\\
	&= -\Big[\lambda_{Q_L}\left(\tilde{Q}^*_{u_L}\bar{\chi} u_L+\tilde{Q}^*_{d_L}\bar{\chi} d_L\right) + \lambda_{u_R}\tilde{Q}^*_{u_R}\bar{\chi} u_R + \lambda_{d_R}\tilde{Q}^*_{d_R}\bar{\chi} d_R + \mathrm{h.c.}\Big]\,,
\end{align}
with $\tilde{Q}_L = (\tilde{Q}_{u_L}, \tilde{Q}_{d_L})^T$ and $Q_L = (u_L, d_L)^T$ denoting SU(2)$_L \times\allowbreak $U(1)$_Y$ doublets. Here,  the $\tilde{Q}_{u_L}$, $\tilde{Q}_{d_L}$, $\tilde{Q}_{u_R}$, $\tilde{Q}_{d_R}$ are the scalar mediator fields with a common mass $\MQ$, and the $u_{L/R}$, $d_{L/R}$ are the left- and right-handed up- and down-type quark fields. Flavor indices are suppressed and we generically  write ``$u$'' for all up-type quarks and ``$d$'' for down-type quarks. The $\lambda_{Q_L}$, $\lambda_{u_R}$, and $\lambda_{d_R}$ are the Yukawa couplings of the mediator to the left- and right-handed quark fields, respectively. Due to the SU(2)$_L\times$U(1)$_Y$ symmetry of the mediators, the left-handed quark coupling $\lambda_{Q_L}$ is identical for up- and down-type flavors. Note that while in the $s$-channel model the mediator can be lighter than the DM particle, $t$-channel mediators are always heavier than the DM particles for them to remain stable. In the following, we will denote by $\Mmed$ the mediator mass of the corresponding simplified model.

%%%%%%%%%%
\subsection{Direct detection in the non-relativistic limit}
\label{s:theory:dd}
The experimental signature of a DM particle of mass $m_\chi$ scattering elastically off a nuclear target particle of type $i$ and mass $m_i$ is given by the differential rate of nuclear recoil events $\totd R$ per energy interval $\totd E$,
\begin{align}
\label{eq:dd-rate}
	\frac{\totd R}{\totd E} = \sum_{i} k_i\frac{\sigma_i}{2m_\chi\mu_i^2}\rho_0\eta_i\,,
\end{align}
where $k_i$ denotes the mass fraction of nuclear species $i$ in the detector, $\sigma_i$ the DM-nucleus scattering cross section,  $\rho_0$ the local DM density,  and $\mu_i$ the reduced mass of the DM-nucleus system,
\begin{align}
	\mu_i = \frac{m_\chi m_i}{m_\chi + m_i}\,.
\end{align}
The quantity $\eta_i$ depends on the distribution of the velocity $v$ of the DM particles relative to the detector,
\begin{align}
\eta_i = \int_{v_{\mr{min},i}}^{v_\mr{esc}}
\totd^3 v \frac{f(\vec{v})}{v}
\quad\text{with}\quad
v_{\mr{min},i}=
\sqrt{\frac{m_i E}{2\mu_i^2}}\,,
\end{align}
where the integration range is determined by the minimum velocity $v_{\mr{min},i}$ of the DM particle required to cause a recoil energy $E$ in the detector, and the galactic escape speed $v_\mr{esc}$ beyond which DM particles are no longer gravitationally bound in the Milky Way. The quantity $f(\vec{v})$ encodes the local velocity distribution in the detector rest frame.

DM particles move at non-relativistic velocities of about $v \sim 10^{-3} c$~\cite{Lin:2019uvt}. Assuming the nuclei in the detector to be at rest, typical recoil energies, which are proportional to the momentum exchange $q$ between the DM particles and the nuclei, are of the order of a few to a few hundred keV. We will therefore work in the limit of vanishing relative momentum between scattering DM particle and nucleus, where $q^2 \to 0$, and refer to it as the \emph{non-relativistic limit}.

In the differential detection rate of Eq.~(\ref{eq:dd-rate}) all information on the microscopic DM-nucleus interaction is contained in the elastic DM-nucleus cross section $\sigma_i$, while the other quantities entering the equation are related to the detector composition and the DM relic density. In the non-relativistic limit, this elastic scattering cross section receives contributions only from so-called spin-independent (SI) and spin-dependent (SD) interactions. The cross section $\sigma_i$ therefore can be divided into an SI ($\sigma^\si_i$) and an SD ($\sigma^\sd_i$) part. The strength of the contribution will differ for different types of nuclei, as the SI part is enhanced for heavy nuclei, while this enhancement is not present for the SD part (see, e.g., Ref.~\cite{Schumann:2019eaa}). In order to allow for easier comparisons between different experimental setups using different detector materials, results are usually expressed in terms of the cross sections of the scattering between a single nucleon and a DM particle, keeping the division between the SI ($\sigma^\si_N$) and SD ($\sigma^\sd_N$) parts.

At low energies, elastic DM-nucleon scattering can be described by an effective four-fermion interaction which can be parameterized in terms of effective operators. We therefore start our discussion with an effective Lagrangian in relativistic notation with DM-quark interaction terms of the form
\begin{align}
	\mc{L}_\mr{EFT}^{\mr{int}, \chi q} &= c_{S}\, \mathcal{O}_S + c_{V}\, \mathcal{O}_V + c_{A}\, \mathcal{O}_A + c_{T}\, \mathcal{O}_T\notag\\
	&= c_{S} \big[\bar{\chi}\chi\, \bar q q\big] + c_{V} \big[\bar{\chi} \gamma^\mu \chi \,\bar{q} \gamma_\mu q\big] + c_{A} \big[\bar{\chi} \gamma^\mu\gamma_5\, \chi \bar{q} \gamma_\mu\gamma_5 q\big] - \frac{1}{2} c_{T} \big[\bar{\chi} \sigma^{\mu\nu} \chi \, \bar{q} \sigma_{\mu\nu} q\big]\,,\label{eq:leff}
\end{align}
with $\sigma_{\mu\nu} = \frac{i}{2}[\gamma_\mu, \gamma_\nu]$, and a priori unknown Wilson coefficients $c_j \equiv c_{j,q}$ which in general depend on the quark flavor $q$. This Lagrangian contains the effective dimension-six operators $\mathcal{O}_j$ accounting for scalar, vector, axial-vector, and tensor interactions of four fermions. These are the only operators that are not suppressed by small non-relativistic velocities. Mixed operators, such as e.g.\ $\bar{\chi}\gamma^\mu\chi\, \bar q \gamma_\mu\gamma_5 q$, are kinematically suppressed and will therefore not be discussed further. In the non-relativistic limit, the scalar and vector operators induce SI, the axial-vector and tensor operators SD interactions.

In order to derive from the EFT Lagrangian the scattering cross sections relevant for direct detection experiments, we can compute for each term in the Lagrangian the matrix element
\beq
	\label{eq:matrixelement-eft}
	\mc{M}_{j,q} = c_{j,q} \, \left(\bar{\chi} \Gamma_j \chi\right) \, \left(\bar{q} \Gamma_j q\right)\,,
\eeq
where for $j=S,V,A,T$ the $\Gamma_j$ stands for $1, \gamma_\mu, \gamma_\mu\gamma_5, \frac{i}{\sqrt{2}}\sigma_{\mu\nu}$, respectively. For computing quantities involving nucleons rather than elementary quarks, we adopt the conventional assumption that, in the non-relativistic limit, quark operators within nucleonic states are proportional to nucleonic operators. This effectively leads to a replacement of the quark fields by nucleon fields (see, e.g., Refs.~\cite{Dienes:2013xya}),
\beq
	\bar{q} \,\Gamma_j q \quad\longrightarrow\quad \sum_q f_{j,q}^N\, \bar{N} \Gamma_j N\,,
\eeq
with the nucleonic matrix elements $f_{j,q}^N$ as proportionality factors, which encode the non-perturbative contributions from hadronic physics and are typically calculated in lattice-gauge theory or determined experimentally. The $f_{j,q}^N$ depend on the type of the nucleon $N = p, n$, the interactions $j$, and the quark flavor $q$.
\begin{table}
	\centering
	\begin{tabular}{c|cccc}
		\toprule
		& \multicolumn{4}{c}{quark flavor}\\
		& $d$ & $u$ & $s$ & $c$, $b$, $t$\\
		\midrule
		$f_{S,q}^p$ ($f_q^p$) & $0.0191$ & $0.0153$ & $0.0447$ & $0.0682$\\
		$f_{S,q}^n$ ($f_q^n$) & $0.0273$ & $0.0110$ & $0.0447$ & $0.0679$\\
		$f_{V,q}^p$ ($f_{V_q}^p$) & $1$ & $2$ & $0$ & $0$\\
		$f_{A,q}^p$ ($\Delta_q^p$) & $-0.427$ & $0.842$ & $-0.085$ & $0$\\
		$f_{T,q}^p$ ($\delta_q^p$) & $-0.230$ & $0.840$ & $-0.046$ & $0$\\
		\bottomrule
	\end{tabular}
	\caption{Numerical values of the nucleonic matrix elements used in this work. Indicated in brackets are alternative names of the coefficients commonly used in the literature.
	\label{tbl:nucleonfactors}}
\end{table}
In our calculations we use the numerical values for the $f_{j,q}^N$ listed in Tab.~\ref{tbl:nucleonfactors}. They have been obtained with the program \micromegas{} \cite{Belanger:2008sj,Belanger:2018ccd}, version 5.0.9. Note that values of the vector, axial-vector, and tensor coefficients for neutrons can be obtained from the respective proton coefficients using isospin symmetry, i.e.\ by switching the $d$ and $u$ columns. 

In order to capture all the quark content of the nucleon, a summation over all active quark flavors has to be performed. We therefore introduce the nucleonic couplings
\begin{equation}
	\label{eq:nucleoncouplings}
	g_j^N = \sum_q f_{j,q}^N c_{j,q}
\end{equation}
for each interaction type $j$. The total cross section for nucleon scattering in the non-relativistic limit then simply reads:
\beq
	\sigma_N = \frac{1}{16 \pi s} \overline{\sum_{\text{pol.}}}\left|\mc{M}_N\right|^2\qquad\text{with } \mc{M}_N \equiv \sum_j \mc{M}_{j,N}\,,
\eeq
where the matrix element $\mc{M}_{j,N}$ is obtained from $\mc{M}_{j,q}$ by replacing the quark currents and Wilson coefficients with the corresponding nucleonic currents and couplings, and averaging (summing) over initial-state (final-state) polarizations and colors has been taken into account as indicated by the bar. In the non-relativistic limit, the center-of-mass energy squared of the DM-nucleon system can be written as $s = \left(\mchi + m_N\right)^2$, with $m_N$ being the mass of the considered nucleon.

Distinguishing between the contributions to SI and SD scattering and evaluating the fermionic traces within the squared amplitude, we obtain 
\begin{align}
	\label{eq:xsec-sisd}
	\begin{split}
		\sigma^\si_N &= \frac{ \mu_N^2 }{\pi} \left| g_S^N \pm g_V^N \right|^2 = \frac{ \mu_N^2 }{\pi} \left| g_\si^N \right|^2\,,\\
		\sigma^\sd_N &= \frac{3 \mu_N^2 }{\pi} \left| g_A^N \pm g_T^N \right|^2 = \frac{3 \mu_N^2 }{\pi} \left| g_\sd^N \right|^2\,,
	\end{split}
\end{align}
with the reduced mass $\mu_N$ of the DM-nucleon particle system. The positive (negative) sign between the contributions from the different Wilson coefficients corresponds to DM (anti-DM) scattering.

In this work, we consider DM scattering in the framework of the simplified models introduced in Sec.~\ref{s:theory:models}. However, experimental results for direct detection experiments are often quoted in terms of EFT quantities. We therefore need to establish a strategy for a translation between the two approaches. We closely follow the procedure of Ref.~\cite{Klasen:2016qyz} where a matching between neutralino-parton scattering in the framework of the MSSM and respective expressions in an EFT approach was presented. Instead of the SUSY Lagrangian of that reference we use the Lagrangian interaction terms of Eqs.~(\ref{eq:L-sim-mod-s})--(\ref{eq:L-sim-mod-t}) for the computation of scattering amplitudes in the framework of simplified models. To find the expressions of the coefficients $c_{j,q}$ in terms of the physical parameters of our models, we match, at the quark level, the simplified model amplitudes to the respective EFT expressions of \refeq{eq:matrixelement-eft}, by imposing, at each order in perturbation theory,
\beq
	\label{eq:matchLO}
	\mc{M}_\mr{sim} \stackrel{!}{=} \mc{M}_\EFT,
\eeq
where $\mc{M}_\mr{sim}$ and $\mc{M}_\EFT$ denote the amplitude for DM-quark scattering in the simplified model and the EFT, respectively. We note that $\mc{M}_\EFT$ is a function of the Wilson coefficients $c_{j,q}$, while $\mc{M}_\mr{sim}$ depends on the parameters of the simplified model. For specific settings of the simplified model parameters then a comparison with experimental limits on the Wilson coefficients can be performed, see, e.g., Ref.~\cite{Angloher:2018fcs} for the CRESST-II experiment.

We note that the simplified model is valid at a high-energy scale, which we choose to be $\mu_{\text{high}} = M_{\text{med}}$, and thus the matching, demanding the EFT to reproduce the full theory, is to be performed at this scale. As the nucleonic matrix elements are, however, defined at a low-energy scale $\mu_{\text{low}} \sim 2$~GeV, the Wilson coefficients $c_{j,q}$ consequently have to be evolved from $\mu_{\text{high}}$ down to $\mu_{\text{low}}$ via renormalization-group equations.

%%%%%%%%%%
\subsection{Radiative corrections in the non-relativistic limit}\label{s:theory:radcor}
%%%%%%%%%%
\begin{figure}[tp]
	\centering
	\addtolength{\tabcolsep}{-3mm}
	\begin{tabular}{cc}
		\includegraphics[scale=.72]{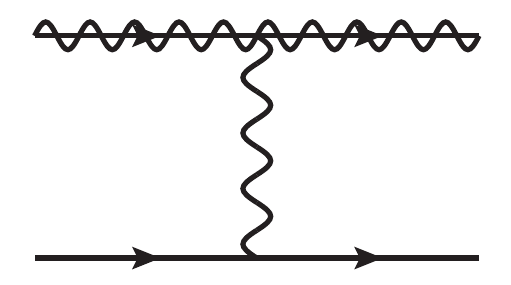} & \includegraphics[scale=.72]{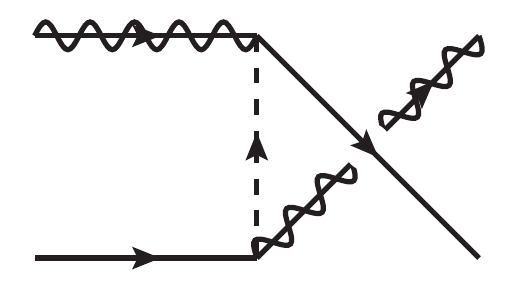}\\
		(a) & (b)
	\end{tabular}
	\addtolength{\tabcolsep}{3mm}
	\caption{Tree-level Feynman diagrams for DM-quark scattering, for the discussed $s$-channel (a) and $t$-channel (b) models.}
	\label{feynman_diags_LO}
\end{figure}
%
%%%%%%
%
\begin{figure}[tp]
	\centering
	\addtolength{\tabcolsep}{-3mm}
	\begin{tabular}{cc}
		\includegraphics[scale=.72]{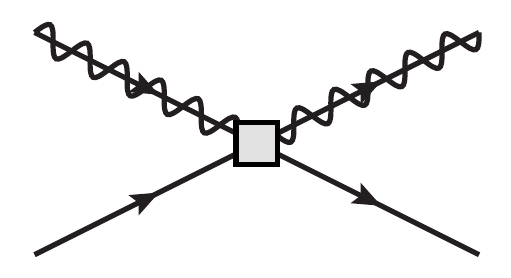} & \includegraphics[scale=.72]{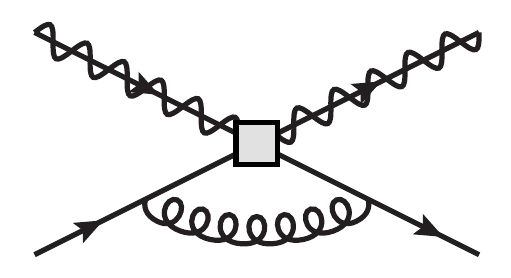}\\
		(a) & (b)
	\end{tabular}
	\addtolength{\tabcolsep}{3mm}
	\caption{Tree-level (a) and one-loop (b)  Feynman diagrams for DM-quark scattering in the effective theory, where the four-fermion interaction is indicated by a grey square.
	}
	\label{feynman_diags_EFT}
\end{figure}
%
%%%%%%
%
The leading-order (LO) Wilson coefficients contributing to SI and SD scattering are determined by matching the tree-level amplitudes of our simplified models, shown in \reffig{feynman_diags_LO} for the $s$- and $t$-channel models, respectively, to the lowest-order operator expression in the EFT, see \reffig{feynman_diags_EFT}~(a). As the tree-level reactions do not involve any strong interactions, the associated cross sections are independent of the strong coupling $\alpha_s$ and depend purely on the model-specific couplings between the SM and DM sectors.

\begin{figure}[tp]
	\centering
	\addtolength{\tabcolsep}{-3mm}
	\begin{tabular}{cccc}
		\includegraphics[scale=.72]{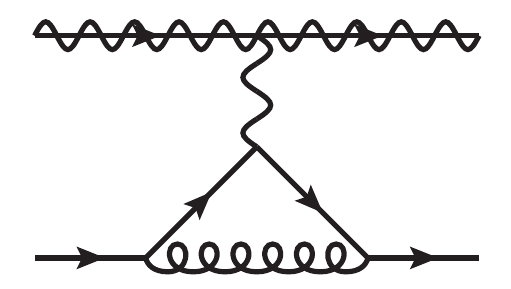} & \includegraphics[scale=.72]{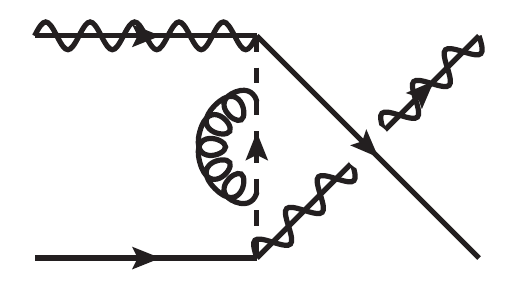} & \includegraphics[scale=.72]{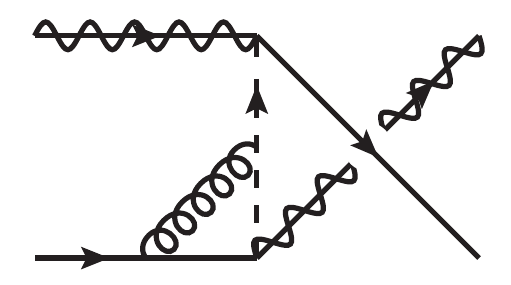} & \includegraphics[scale=.72]{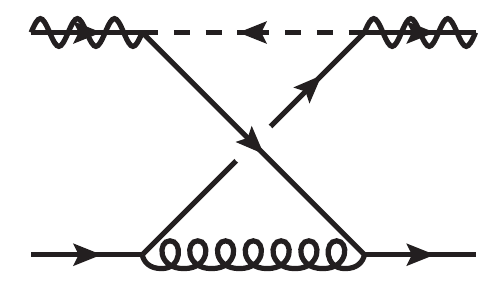}\\
		(a) & (b) & (c) & (d)
	\end{tabular}
	\addtolength{\tabcolsep}{3mm}
	\caption{Representative one-loop Feynman diagrams for DM-quark scattering, for the $s$-channel~(a) and $t$-channel [(b)--(d)] models.}
	\label{feynman_diags_NLO}
\end{figure}
In order to improve the tree-level predictions for the Wilson coefficients, all the one-loop diagrams in the considered simplified models that contribute to the relevant operators at $\mathcal{O}(\alphas)$ have to be calculated. In \reffig{feynman_diags_NLO} representative one-loop diagrams in the $s$- and $t$-channel models are depicted. They can be classified as  propagator, vertex, and box corrections, and have to be considered together with the corresponding counterterm diagrams. We denote these additional contributions as NLO-QCD or simply NLO corrections, as in this work we consider a perturbative expansion only in the QCD coupling $\alphas$.

Loop-induced processes with gluons in initial and final states contributing to DM-nucleon scattering have been calculated in the framework of the minimally supersymmetric extension of the SM and similar models in \cite{Drees:1992rr,Drees:1993bu,Hisano:2010ct,Hisano:2015bma,Hisano:2015rsa}. We want to clarify that formally these contributions contribute only at $\mathcal{O}(\alphas^2)$ to the cross sections, which is one order higher than what we are discussing in our work, and we are thus not taking them into account.

Let us remark that while in general a full NLO-QCD calculation involves both virtual and real-emission corrections, in the non-relativistic limit real-emission contributions do not have to be considered. In this limit the in- and outgoing DM and quark momenta become equal, and no extra parton emission occurs. Consequently, the sum of all virtual corrections is infrared finite, which we have verified explicitly.

In order to convert the NLO corrections in the context of the simplified models we consider to one-loop expressions for the Wilson coefficients of the EFT, the matching condition of \refeq{eq:matchLO} has to be applied at NLO. Additionally, the one-loop operator corrections sketched in \reffig{feynman_diags_EFT}~(b) have to be considered. Spelled out in terms of effective operators, the NLO matching condition reads
\beq
\label{eq:matchNLO}
	\mc{M}_\mr{sim}^\nlo \stackrel{!}{=} \sum_{j = S, V, A, T} c_j^\nlo \mc{O}_j^\nlo\,.
\eeq
It is important to note that in this relation both the Wilson coefficients and the effective operators are subject to NLO corrections. The full matrix element $\mc{M}_\mr{sim}^\nlo$ as well as each term in the sum, $c_j^\nlo \mc{O}_j^\nlo$, can be decomposed into tree-level and one-loop parts,
\[
	\mc{M}_\mr{sim}^\tree + \mc{M}_\mr{sim}^\oneloop ~\text{ and }~ \left(c_j^\tree + c_j^\oneloop\right) \left(\mc{O}_j^\tree + \mc{O}_j^\oneloop\right),
\]
respectively.

In the latter we do not compute terms like $c_j^\oneloop\mc{O}_j^\oneloop$, as they contribute at $\mathcal{O}(\alphas^2)$, which is beyond the perturbative accuracy we are interested in. In the simplified $s$- and $t$-channel models that we consider, $c_S^\tree = c_T^\tree = 0$ (see \ref{s:NLO-expressions} for details). At tree level we therefore obtain the following matching condition:
\begin{equation}
	\label{eq:wil-coeff-tree}
	\mc{M}_\mr{sim}^\tree \stackrel{!}{=} c_A^\tree \mc{O}_A^\tree + c_V^\tree \mc{O}_V^\tree\,.
\end{equation}
At $\mathcal{O}(\alphas)$, after inserting \refeq{eq:wil-coeff-tree}, we find: 
\begin{align}
	\label{eq:wil-coeff-oneloop}
	\begin{split}
		&\mc{M}_\mr{sim}^\oneloop - c_A^\tree \mc{O}_A^\oneloop - c_V^\tree \mc{O}_V^\oneloop \stackrel{!}{=}\\
		&\qquad~ c_S^\oneloop \mc{O}_S^\tree + c_A^\oneloop \mc{O}_A^\tree + c_V^\oneloop \mc{O}_V^\tree + c_T^\oneloop \mc{O}_T^\tree\,.
	\end{split}
\end{align}
Calculating the quantities appearing on the left-hand side of this equation, $\mc{M}_\mr{sim}^{\oneloop}$ and $\mc{O}_j^\oneloop$, we are able to identify the individual Wilson coefficients on the right-hand side by comparing the related  tensor structures. For instance, contributions to the scalar Wilson coefficient $c_S^{\oneloop}$ stem from the fermion chain $\bar{\chi}\chi\, \bar q q$ in the matrix element $\mc{M}_\mr{sim}^{\oneloop}$. Details of the calculation as well as explicit expressions for the one-loop Wilson coefficients and operator corrections are given in \ref{s:NLO-expressions}.

An interesting feature of the $t$-channel model is that in this scenario at tree level only the Wilson coefficients of the vector and axial-vector operators, $c_V$ and $c_A$, contribute. Genuine contributions to the scalar and tensor coefficients, $c_S$ and $c_T$, are obtained only starting at order $\mc{O}(\alpha_s)$. However, we have examined the numerical effects of all Wilson coefficients and found that, compared to the dominant Wilson coefficients, the contribution of $c_S$ and $c_T$ to the full NLO cross section is small: it is less than 0.1\% in most cases, and in the range of 0.1\% up to approximately 1\% for low DM masses and when being close to the threshold of equal DM and mediator masses.

%%%%%%%%%%%%%%%%%%%%%%%%%%%%%%%%%%%%%%%%%%%
\section{Numerical analysis}\label{s:results}
%%%%%%%%%%%%%%%%%%%%%%%%%%%%%%%%%%%%%%%%%%%
%
We now present numerical results for the SI and SD DM-nucleon scattering cross sections, and discuss the effects of the $\mathcal{O}(\alphas)$ corrections. Furthermore, we assess the potential of various direct detection experiments to explore regions in their respective parameter spaces when confronted with limits of LHC searches for DM production. As the numerical difference between the proton and neutron contribution to the SI cross section is marginal and the direct detection experiments typically publish limits irrespective of the nature of the nucleon, we limit ourselves to showing only the DM-proton cross sections for SI interactions. On the other hand, since the sensitivities of the experiments for the SD interaction differ between proton and neutron scattering, and the CRESST collaboration has only published limits for neutron scattering, we will focus on the DM-neutron cross sections in the case of SD interactions.

In the following, we do not discuss the pure $s$-channel model with a vector mediator further, since the $\mathcal{O}(\alphas)$ corrections vanish for such a scenario, as discussed in \ref{s:app:schannel-corrections}. The impact of $s$-channel contributions is only considered in the context of interference effects in models that feature $s$- and $t$-channel topologies at the same time (see also \ref{s:app:stchannel-corrections}).
For the $t$-channel model, we assume the DM particle to be a Dirac fermion. We choose a scenario where all right-handed couplings are turned off, i.e.\ $\lambda_{u_R} = \lambda_{d_R} = 0$ for all quark flavors, and the mediators thus only couple to the left-handed SU(2)$_L$ quark doublets. This choice is motivated by the most recent mono-jet searches of the ATLAS experiment \cite{Aaboud:2017phn}, where limits are presented for the same selection of couplings. In the following, we use the short-hand notation $\lambda \equiv \lambda_{Q_L}$, and assume the mediators to couple to all left-handed quark doublets with equal strength. Unless explicitly specified otherwise, we choose $\lambda = 1$.

All quarks in our calculations are assumed to be massive and we use for the quark masses the current values compiled by the particle data group~\cite{Zyla:2020abc}. However, as in our models the dominant contributions for the SI and SD scattering processes stem from the vector and axial-vector operators, which receive non-vanishing contributions only from the light $u$, $d$, and $s$ quark flavors, the numerical effects of the light quark masses are negligibly small. For the nuclear matrix elements we use the numerical values quoted in Tab.~\ref{tbl:nucleonfactors}.

We use the world average for the strong coupling evaluated at the mass of the $Z$~boson, $\alphas(M_Z) = 0.1179$ \cite{Zyla:2020abc}, as an input value for determining the value of $\alphas(\mur)$ at a value of the renormalization scale $\mur$ representative for the considered class of reactions via NLO-QCD running of the renormalization group equation.
Since DM scattering reactions at direct detection experiments typically occur at rather low scales, we choose $\mur = 2$~GeV as our default value and discuss uncertainties arising from a variation of $\mur$ by a factor of two around the default value, leading to a variation of the value of $\alphas$ in the range from 0.226 to 0.443. We note that for such low scales, the strong coupling assumes rather large values, in particular $\alphas(\mur=1~\mr{GeV}) = 0.443$. Thus, contrary to the situation at high-energy colliders such as the LHC, perturbative calculations in the context of direct detection experiments have to be taken {\em cum grano salis}, and the potential uncertainties associated with perturbative predictions have to be carefully assessed. We note that an assessment of non-perturbative contributions, such as higher-twist corrections or nuclear structure effects, is not covered by our approach and would require complementary means.

%%%%%%%
Let us briefly comment on the dependence of the NLO SI and SD cross sections on the renormalization scale $\mur$. Because of the particular structure of the reactions we consider in the framework of our simplified models, $\alphas(\mur)$ enters first in the NLO-QCD corrections, while the hard parts of the LO cross sections are independent of $\alphas$ and of $\mur$. Schematically, the perturbative expansion of these cross sections, generically denoted by $\sigma$, is of the form
\bea
	\sigma = \sigma^{(0)} + \alphas(\mur) \sigma^{(1)}(\mur) + \mc{O}(\alphas^2)\,,
\eea
with all dependence on $\alphas$ explicitly factored out of the expansion coefficients $\sigma^{(i)}$. Varying $\mur$ thus affects both, the value of $\alphas$, and the size of the NLO correction term $\sigma^{(1)}$ which depends explicitly on renormalization scale logarithms. In the following discussion of our results, we will study the effects of varying the renormalization scale.

The above discussion does not yet include any renor\-mal\-iza\-tion-group running of the effective operators and, consequently, of the Wilson coefficients. As discussed in the previous section, the Wilson coefficients, having been determined at a high scale $\mu_{\text{high}}$, have to be evolved down to the scale $\mu_{\text{low}}$ at which the nucleonic matrix elements are evaluated. Then, as the running effects in the product of Wilson coefficient and nucleonic matrix element are inversely proportional and therefore cancel, the identification of $\mu_{\text{low}}$ with the value of the renormalization scale $\mur$ and a subsequent variation of $\mur$ will not lead to an additional source of scale uncertainty%
\footnote{In principle, $\mu_{\text{high}}$ could also be varied, leading to an additional uncertainty related to the renormalization-group running of the Wilson coefficients. We have checked that the effect of this variation is small compared to the other scale uncertainties in our work and we therefore do not discuss it further.}%
. We take into account the running effects of the operators appearing in our models following Ref.~\cite{Hill:2014yxa}, i.e.\ we include the effects of the resummation of QCD logarithms only%
\footnote{A general way to include renormalization-group-running effects, also comprising the resummation of electroweak/QED logarithms, can be achieved, for instance, by using tools such as \cite{Bishara:2017nnn,Fitzpatrick:2012ix}.}%
. In practice, we find that the running is numerically only relevant for the axial-vector operator, though, for which we apply the two-loop running coefficient: while there are no running effects for the vector operator, the contributions of the scalar and tensor Wilson coefficients are too small for any running effects to be significant.

Using the specified setup, we first discuss the impact of NLO-QCD corrections and scale uncertainties on SI and SD scattering cross sections. After this assessment of the genuine features of experimentally accessible observables we turn to a systematic comparison of limits on DM models from colliders and direct detection experiments in the context of $t$-channel models. Subsequently, we investigate the impact of interference effects in combined $s+t$-channel models.

%%%%%%%%%%%%%%%%%%%%%%%%%%%%%%%%%%%%%%%%%%%
\subsection{Theoretical aspects of the perturbative prediction}
\label{s:results:ddnlomu}
%%%%%%%%%%%%%%%%%%%%%%%%%%%%%%%%%%%%%%%%%%%
%
%%%%%%%%%%
%
\begin{figure}[tp]
	\centering
	\includegraphics[width=\textwidth]{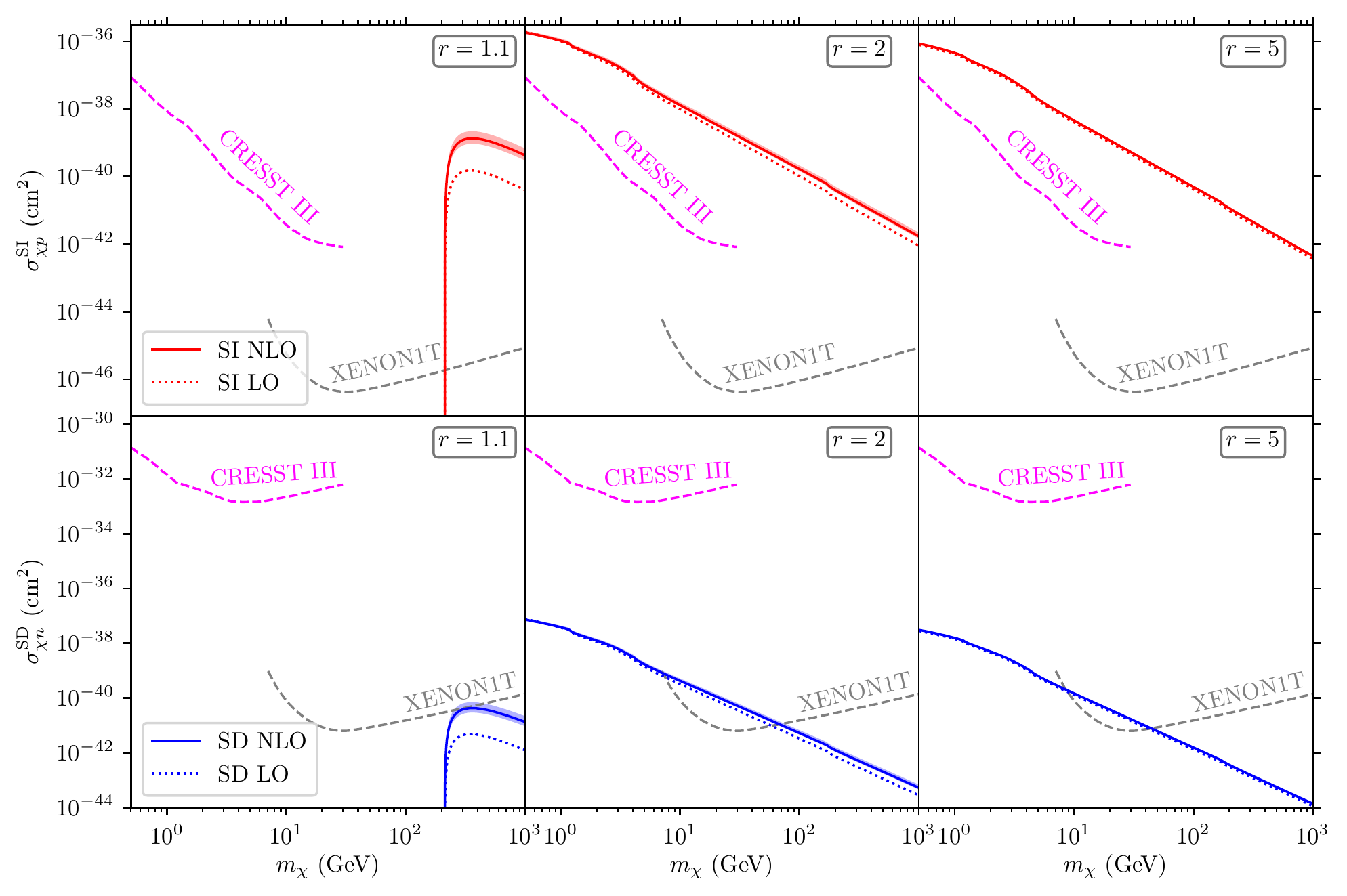}
	\caption{SI DM-proton (top) and SD DM-neutron (bottom) scattering cross sections for the $t$-channel model of \refeq{eq:L-sim-mod-t} for given values of $r = \Mmed/\mchi$ at LO (dotted) and NLO (solid), with bands indicating their renormalization scale dependence. Additionally, current exclusion limits from CRESST \cite{Abdelhameed:2019hmk} (dashed magenta) and XENON \cite{Aprile:2018dbl,Aprile:2019dbj} (dashed grey) are shown. The setting of the coupling $\lambda$ is discussed in the text.}
	\label{sigma_SI_p_SD_n_Tchan}
\end{figure}
%
%%%%%%%%%%
%
Before turning to a phenomenological discussion of the considered DM models, we explore the genuine features and theoretical uncertainties of our perturbative calculation.

In \reffig{sigma_SI_p_SD_n_Tchan} we show the SI and SD DM-nucleon scattering cross sections for the $t$-channel model of Eq.~(\ref{eq:L-sim-mod-t}) as functions of the DM mass for fixed values of the ratio  $r= \Mmed/\mchi$, together with curves indicating that the entire region of parameter space above is excluded by the CRESST III \cite{Abdelhameed:2019hmk} and XENON1T \cite{Aprile:2018dbl,Aprile:2019dbj} experiments.
%
%%%%%%%%%%
%
\begin{figure}[tp]
	\centering
	\includegraphics[width=.47\textwidth]{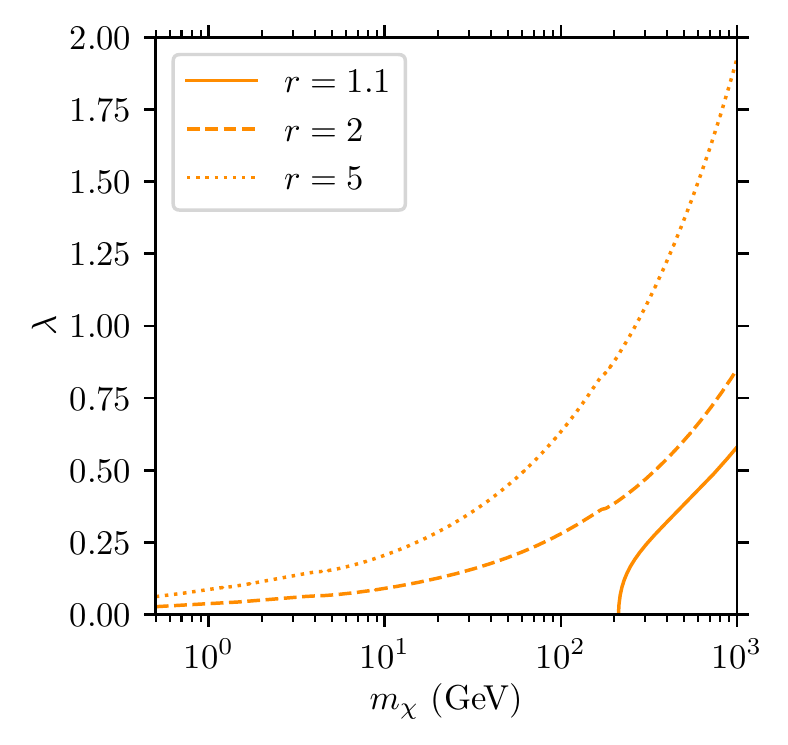}
	\caption{Value of the $t$-channel coupling $\lambda$ depending on $\mchi$ and $r = \Mmed/\mchi$ for which the model correctly predicts the measured DM relic abundance $\Omega_{\text{DM}}h^2 = 0.12$ \cite{Zyla:2020abc}. For $r = 1.1$ and $\mchi < 200$~GeV, no phenomenologically viable configurations are found.}
	\label{tchannel_lambda}
\end{figure}
%
%%%%%%%%%%
%
In contrast to our default assumption of $\lambda = 1$, for this figure we used the \micromegas{} package to calculate the DM relic abundance and extract, for the given values of $\mchi$ and $\Mmed$, the value of the coupling $\lambda$ such that for each point along the curve the model also correctly reproduces the measured DM relic abundance, see \reffig{tchannel_lambda}.

We note that for both, $\sisi$ and $\sisd$, the size of the NLO corrections relative to the LO prediction is largest for $r=1.1$, while it decreases with larger values of $r$. The difference between the LO and the NLO curves is not covered by the scale dependence of the NLO prediction that is indicated by a red band in the figure.
For $r = 1.1$, finite values for the cross sections are obtained only for DM masses above $\sim 200$~GeV. Below that range there are no configurations with a value of $\lambda$ being compatible with the measured value of the relic density. This is no longer the case when larger values of $r$ are considered.

\begin{figure}[tp]
	\centering
	\includegraphics[width=.5\textwidth]{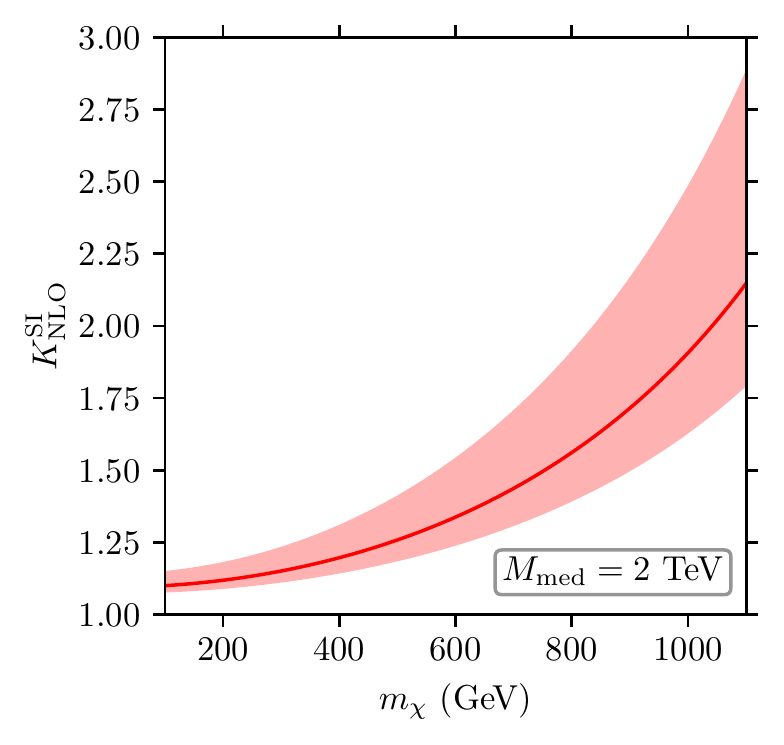}
	\caption{$K$~factor for $\sisi$ in the $t$-channel model of Eq.~(\ref{eq:L-sim-mod-t}) for a fixed mediator mass $\Mmed = 2$~TeV, together with bands indicating the renormalization scale uncertainty. Note that the $K$~factor of $\sisd$ is not shown, as it is numerically very similar to the one for $\sisi$, with differences of at most 1-2\%.}
	\label{kfactor}
\end{figure}

\Reffig{kfactor}
%
%%%%%%%
%
%
%%%%%%%
%
depicts the $K$~factor, defined as the ratio of a NLO-QCD cross section to the respective LO cross section,
\beq
	K_\nlo = \frac{\sigma_\nlo}{\sigma_\mr{LO}}\,,
\eeq
in the spin-independent case, i.e.\ for $\sisi$, as a function of the DM mass.
The value of the coupling $\lambda$ is now set to 1, and for the renormalization scale $\mur$ we choose the central value of 2~GeV. The mediator mass $\Mmed$ is fixed to 2~TeV. In order to illustrate the dependence of $\sisi$ on $\mur$, the scale is varied by a factor of two around this central value. The width of the scale uncertainty band clearly indicates the strong dependence of the NLO prediction on this artificial scale.
Would one instead consider only the LO approximation, being independent of $\mur$, an assessment of the perturbative uncertainty of the prediction would not be possible, and genuine uncertainties would be strongly underestimated.

The considered $K$~factor varies from a value of approximately 1.1 for low $\mchi$ to a value in the range of 1.8 to 2.9 beyond $\mchi = 1$ TeV. It can also be seen that the scale uncertainty grows significantly the closer $\mchi$ is to the threshold $\mchi = \Mmed$. Interestingly, we find that the $K$~factor behaves in a nearly identical manner for SI and SD scattering. This feature is due to the similar structure of the NLO corrections for the dominant vector and axial-vector Wilson coefficients.

\begin{figure}[tp]
	\centering
	\includegraphics[width=.5\textwidth]{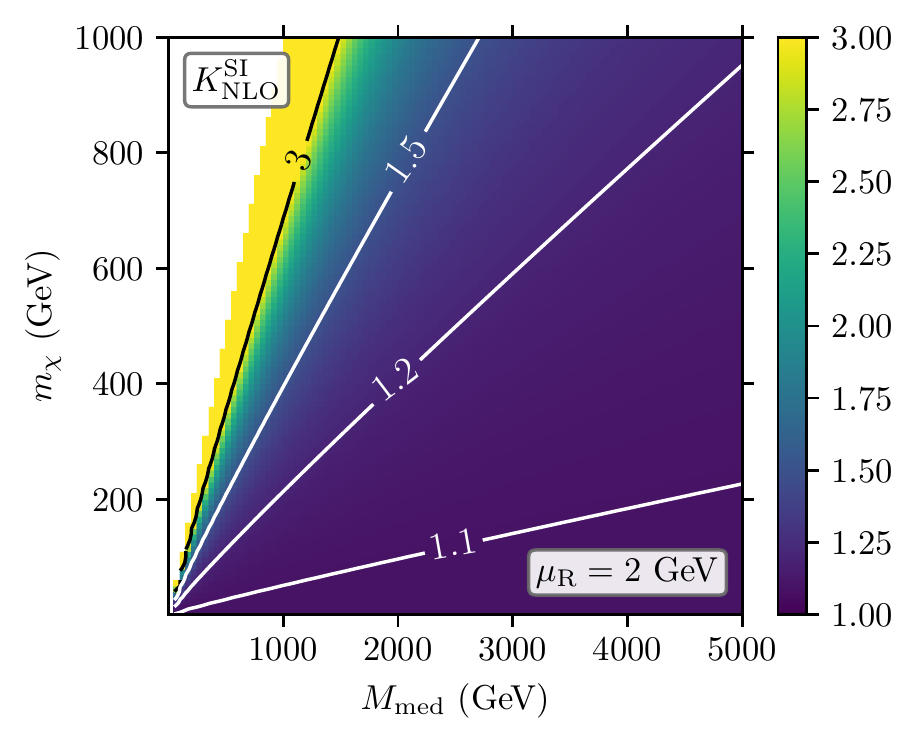}
	\caption{$K$~factor for $\sisi$ for different values of $\Mmed$ and $\mchi$ with a fixed value of $\mur = 2$~GeV.}
	\label{kfactor-fixed-mu}
\end{figure}

To analyze the behavior of the $K$~factor for different mediator masses, \reffig{kfactor-fixed-mu}
%
%%%%%%%
%
%
%%%%%%%
%
illustrates instead the size of the NLO corrections depending on $m_\chi$ and $\Mmed$ for a fixed renormalization scale of $\mur = 2$~GeV in a two-dimensional color-coded plot for $\sisi$. As both the LO and the NLO cross sections diverge at the threshold $\mchi = \Mmed$, $K$~factors larger than three all correspond to a yellow color coding. The white region denotes unphysical parameter points with $\mchi > \Mmed$, where the $t$-channel model does not offer a stable DM candidate. Again, for the same reason as before, we only show the SI case, noting that the SD one looks nearly identical.
The $K$~factor ranges from a value close to one at small $\mchi$ and large $\Mmed$ to large values above two for large $\mchi$ and small $\Mmed$, where, considering the different ranges that we explored, we approach the condition $\mchi = \Mmed$. This means that the ratio between the masses is close to one, and the growth of the $K$~factor in this scenario is in agreement with what we observed in \reffig{sigma_SI_p_SD_n_Tchan}. As expected from our discussion of \reffig{kfactor}, we observe a similar behavior for the SI and the SD case.
%

%%%%%%%%%%%%%%%%%%%%%%%%%%%%%%%%%%%%%%%%%%%
\subsection{Comparison of limits from the LHC and direct detection experiments}\label{s:results:lhcdd}
%%%%%%%%%%%%%%%%%%%%%%%%%%%%%%%%%%%%%%%%%%%
%
We now move on to more phenomenological aspects and discuss the effects of the NLO corrections on the comparison between exclusion limits from LHC searches and direct detection experiments. Let us first discuss some peculiarities of this comparison.

While exclusion limits from collider experiments strongly depend on the studied DM model, they are usually shown in a $M_{\mathrm{med}}$--$m_\chi$ parameter plane for a fixed set of coupling parameters, see e.g.\ \cite{Aaboud:2017phn}. In contrast, direct detection experiments measure the nuclear recoil rate of \refeq{eq:dd-rate}, and the corresponding limits can thus be presented in a model-independent way directly as upper limits on the DM-nucleon scattering cross section, as  functions of $m_\chi$.
When translating limits from one representation to the other, care has to be taken, as there are different assumptions going into the determination of exclusion bounds at colliders and direct detection experiments, respectively, see, e.g., Ref.~\cite{Chala:2015ama,Boveia:2016mrp}. In particular, for collider production of DM, the DM relic abundance is of no relevance. Therefore, in the $M_{\mathrm{med}}$--$m_\chi$ plane showing exclusion limits for a specific model, the calculated relic abundance is typically different at each point. This is not an issue for collider limits in the context of simplified models, since mechanisms unrelated to the considered production mode could be responsible for DM annihilation processes in the early universe that lead to the correct relic abundance as measured today.
In contrast, direct detection experiments measure a scattering rate which depends on the product of the DM-nucleon cross section and the local DM density. Thus, limits from direct detection experiments shown in the DM-nucleon cross section plane assume a fixed relic abundance. When comparing LHC limits to excluded regions from direct detection experiments, these implicit assumptions on the relic abundance have to be taken into account.

The transformation of LHC limits which are given as points in the $(\mchi, \Mmed, \lambda)$ parameter space into limits on DM-nucleon cross sections simply amounts to calculating the SI and SD cross sections for parameter points excluded by the LHC. Thus, the limits will be affected by whether LO or NLO expressions for the cross sections are used. For the translation of published LHC limits into the $\sigma_{p/n}^\mr{SI/SD}-m_\chi$ plane we follow the recommendation of Ref.~\cite{Boveia:2016mrp}. Another feature of the LHC limits is that, as parameter regions of the $t$-channel model up to the threshold $\mchi = \Mmed$ are probed, there is a sharp cutoff for the highest excluded value of $\mchi$. Again, we want to stress that the LHC limits are valid only for the considered DM model, while the limits from direct detection experiments do not rely on any model assumptions.

%
%%%%%%%%
%
\begin{figure}[tp]
	\centering
	\includegraphics[width=.5\textwidth]{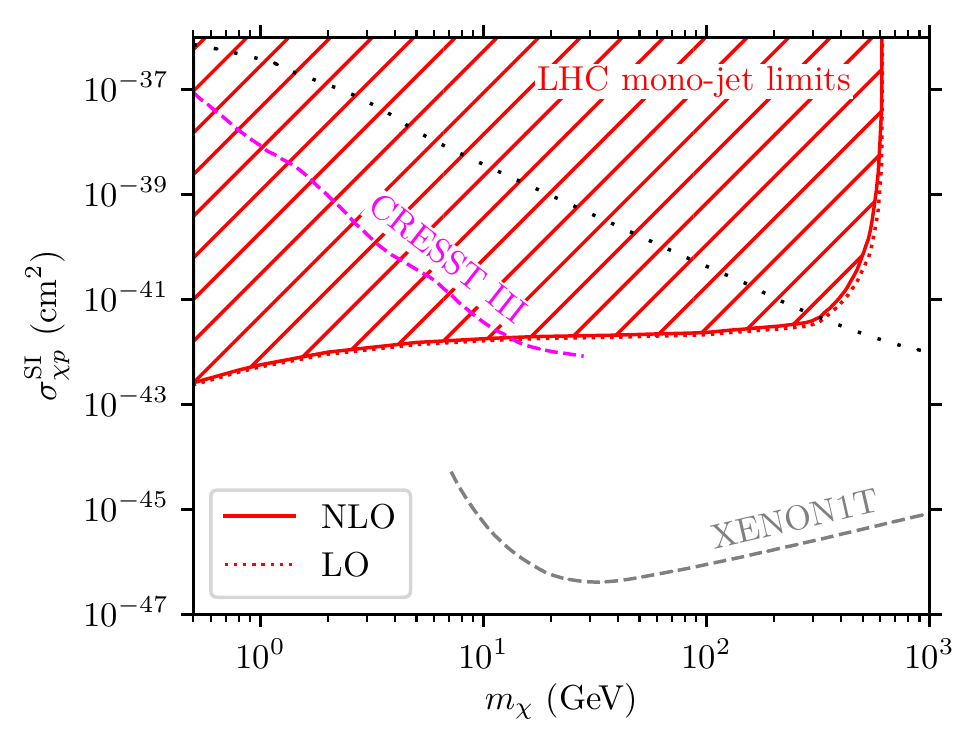}\includegraphics[width=.5\textwidth]{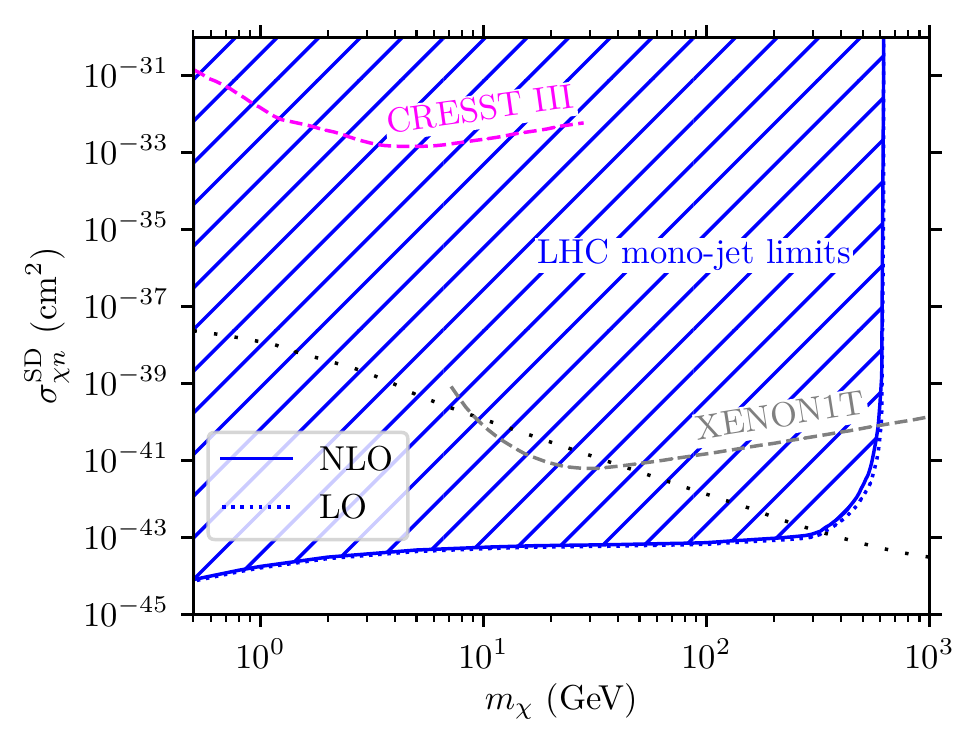}
	\caption{Exclusion limits (hatched areas) for $\sisi$ (left) and $\sisd$ (right), extracted from ATLAS~\cite{Aaboud:2017phn} data at LO and NLO-QCD, and from the CRESST~\cite{Abdelhameed:2019hmk} and XENON~\cite{Aprile:2018dbl,Aprile:2019dbj} experiments for the $t$-channel model of Eq.~(\ref{eq:L-sim-mod-t}), see text for details. Parameter points on the black dotted lines are compatible with the measured relic abundance, the areas above (below) correspond to an under-abundance (over-abundance).}
	\label{compare_LHC_DD_T}
\end{figure}
%
%%%%%%%%%
%
In \reffig{compare_LHC_DD_T} we present exclusion limits for the DM-nucleon scattering cross sections  $\sisi$ and $\sisd$ at LO and NLO accuracy as derived from ATLAS~\cite{Aaboud:2017phn}, together with exclusion limits provided by the CRESST~III~\cite{Abdelhameed:2019hmk} and XENON1T~\cite{Aprile:2018dbl,Aprile:2019dbj} experiments. Numbers are obtained in the context of the $t$-channel model of Eq.~(\ref{eq:L-sim-mod-t}) with its coupling parameter set to $\lambda = 1$. Additionally, we indicate the compatibility of points in the considered parameter space with the measured relic abundance. In the plots, LHC searches for mono-jets as tell-tale signatures of a DM production process yield the hatched red and blue exclusion regions in the two-dimensional plots.
Including NLO-QCD corrections slightly reduces the excluded area as compared to the LO estimate for our default scale choice of $\mur=2$~GeV. Somewhat larger effects would result for a smaller value of $\mur$, as apparent from our discussion of the scale dependence of our calculation in Sec.~\ref{s:results:ddnlomu}. Direct detection limits of the CRESST and XENON experiments are independent of the perturbative corrections. In particular, CRESST excludes the region above the dashed magenta line, while XENON excludes the entire area above the dashed grey line in the plots. The complementary coverage of the two direct detection experiments comes as no surprise, since CRESST is designed to be particularly sensitive to light DM candidates, while XENON performs best in the domain of larger DM masses.
We observe that the considered cross sections diverge for the parameter points of the LHC exclusion limit where the DM mass approaches the mass of the mediator, because of the genuine propagator structure of the scattering amplitude in the $t$-channel model, $\mathcal{M} \sim 1/(m_\chi^2 - M_{\mathrm{med}}^2)$. NLO effects also can be enhanced because of threshold logarithms for $m_\chi \sim \Mmed$. However, generally, the size of the NLO corrections is small, modifying LO results only marginally.

%%%%%%%%%%%%%%%%%%%%%%%%%%%%%%%%%%%%%%%%%%%
\subsection{Phenomenology of a combined $s + t$-channel model}\label{s:results:stchan}
%%%%%%%%%%%%%%%%%%%%%%%%%%%%%%%%%%%%%%%%%%%
Interesting effects can be observed, if instead of a genuine $s$- or $t$-channel model a scenario featuring both types of interactions simultaneously is considered%
\footnote{Arguably, such a model could be considered outside of the domain of the more general class of simplified models, and we discuss it in our work as an example of a more complex DM model.}%
. Conceptually, such a model, termed {\em $s+t$-channel model} in the following, is obtained by adding the two interaction Lagrangians of \refeq{eq:L-sim-mod-s} and \refeq{eq:L-sim-mod-t}. As a consequence, one obtains a model which in addition to the parameters of the genuine $t$-channel model considered previously exhibits the extra parameters of the vector mediator mass $M_V$, the couplings of the (axial) vector mediator to quarks, $g_q^{V(A)}$, and to DM, $g_\chi^{V(A)}$.

In order to assess the features of such a model, we again use $\lambda = 1$ and fix $\Mmed = 500$~GeV, but keep $m_\chi$ as a free parameter, and set the values of the additional parameters in such a way that the pure $s$-channel and $t$-channel contributions are roughly of the same size. In particular, we choose the masses of the two mediator particles and their couplings to the DM particle to be identical, $M_V = \Mmed$, $g_\chi^{V/A} = 1$, and we set $g_q^{V/A} = \pm1/8$. Both values of the sign factor in the coupling of the vector mediator to quarks are considered.

When computing DM-nucleon cross sections in the framework of this $s+t$-channel model, we find that the Wilson coefficients $c^{s+t}$ in principle just correspond to the sum of the Wilson coefficients of the pure $s$- and $t$-channel models, $c^s$ and $c^t$, respectively, as detailed in \ref{s:app:stchannel-corrections}. Thus, the relevant cross sections are sensitive to a relative minus sign between the two contributions and therefore directly to the sign of the $g_q^V$ or $g_q^A$ couplings.
Interestingly, because of the relative minus sign between the two terms at LO, a cancellation occurs which becomes maximal for similar magnitudes of both contributions. At NLO, the cancellation pattern is modified, because of the absence of one-loop corrections in the $s$-channel model, c.f.\ \ref{s:app:schannel-corrections}. Combined with the non-vanishing loop correction to the $t$-channel model, the perturbative expansion of the SI and SD cross sections, generically referred to as $\sigma^\mr{pert}$, is thus, at the perturbative order we consider, schematically of the form 
\begin{equation}
\label{eq:interference-nlo}
	\sigma^\mr{pert} \propto \left|c^{s,\tree} + c^{t,\tree}\right|^2 + 2\operatorname{Re}\Big[\left(c^{s,\tree}+ c^{t,\tree}\right)\left(c^{s + t,\oneloop}\right)^\star\Big] + \left|c^{s + t,\oneloop}\right|^2 + \mathcal{O}(\alphas^2)\,,
\end{equation}
where we have explicitly included the squared one-loop correction which is of order $\mathcal{O}(\alphas^2)$ and thus formally only contributes beyond NLO. Note that we did not include this term in any of the results discussed above. When $\left|c^{s,\tree} + c^{t,\tree}\right|$ is small, the interference term between the tree-level and the one-loop contributions as well as the squared one-loop term can become the dominant contribution to the cross section. If the relative sign between $c^{s,\tree}$ and $c^{t,\tree}$ changes, the hierarchy of terms is reversed, and the one-loop correction ceases to be the dominant contribution to the cross section.

%
%%%%%%%%%
%
\begin{figure}[tp]
	\centering
	\includegraphics[width=\textwidth]{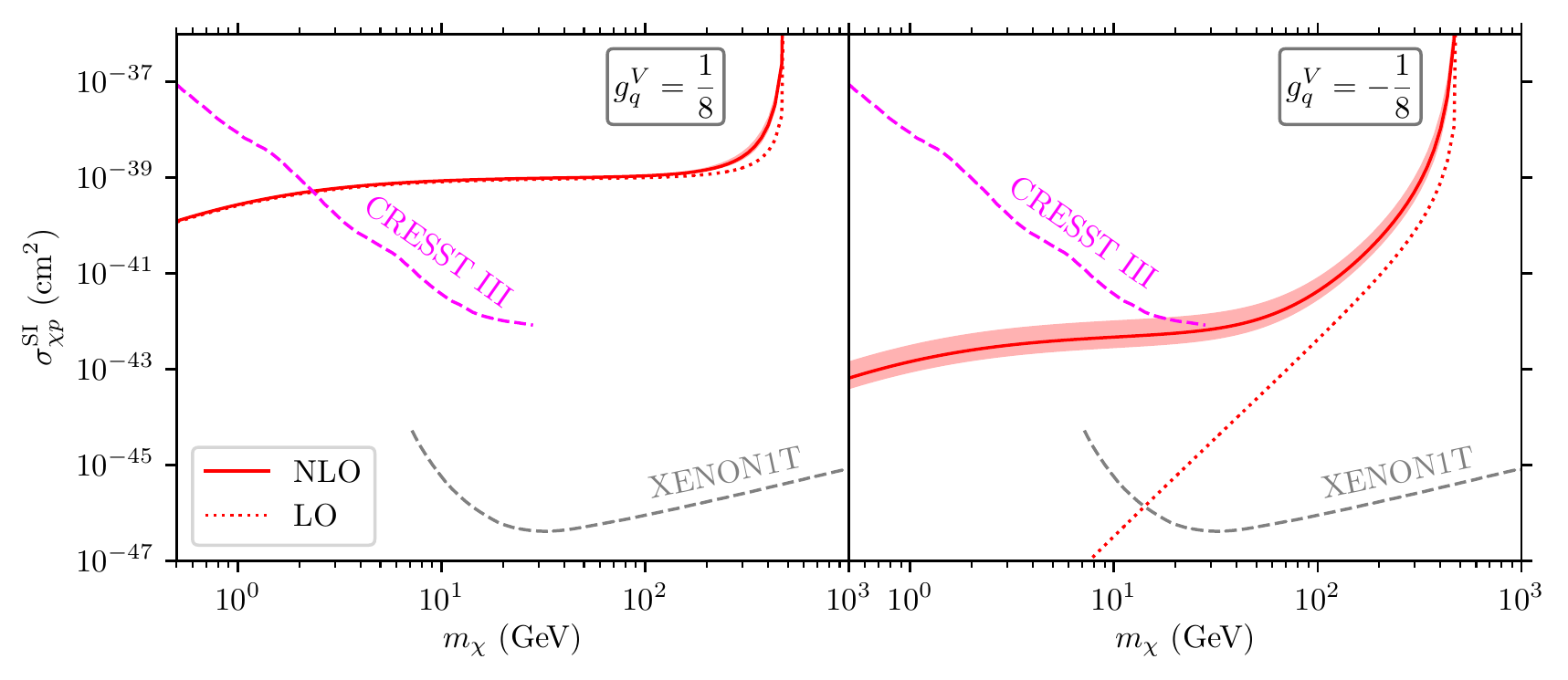}
	\caption{Theoretical predictions at LO (dotted blue) and NLO (solid hatched blue area) for $\sisi$ together with bounds from CRESST and XENON for an $s+t$-channel model with $g^V_q = 1/8$ (left panel) and $g^V_q = -1/8$ (right panel).}
	\label{compare_LHC_DD_SI_ST}
\end{figure}
\begin{figure}[tp]
	\centering
	\includegraphics[width=\textwidth]{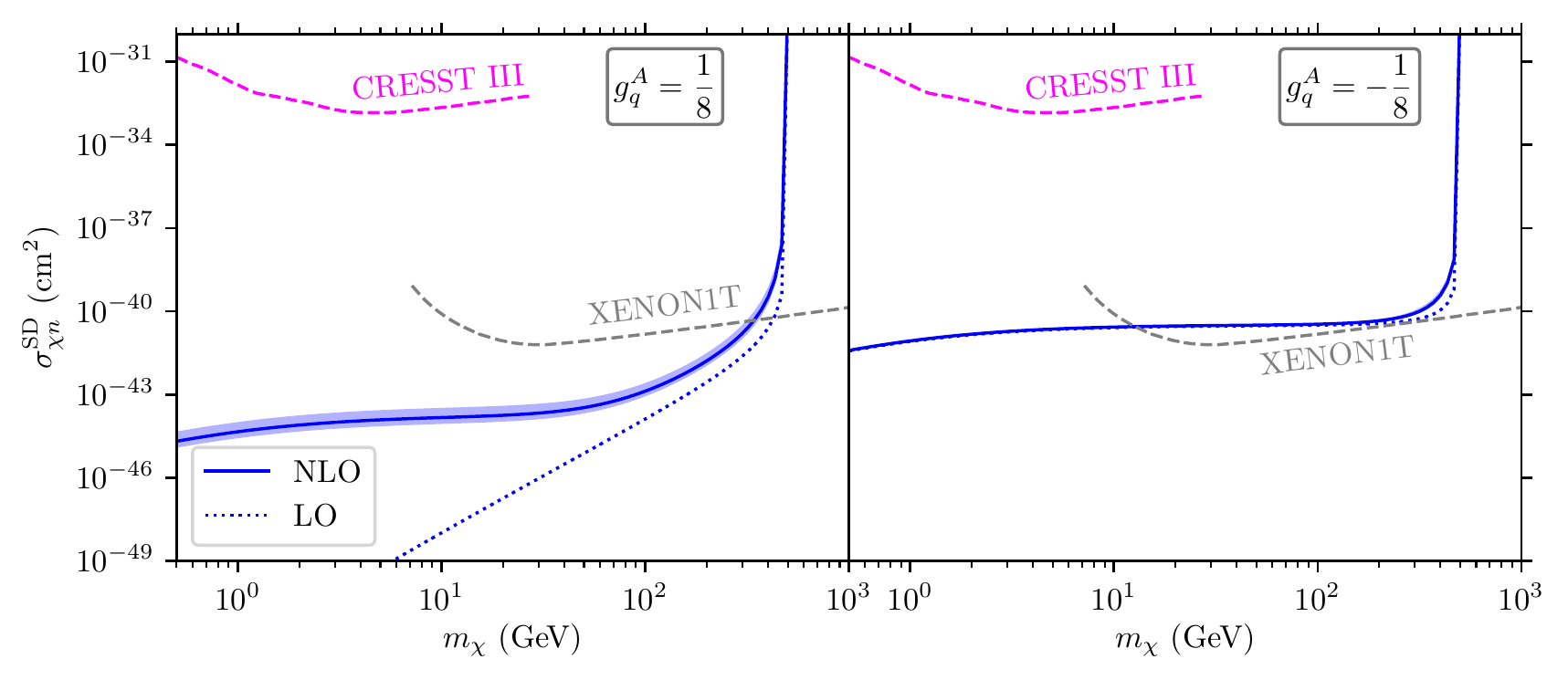}
	\caption{Theoretical predictions at LO (dotted blue) and NLO (solid hatched blue area) for $\sisd$ together with bounds from CRESST and XENON for an $s+t$-channel model with $g^A_q = 1/8$ (left panel) and $g^A_q = -1/8$ (right panel).}
	\label{compare_LHC_DD_SD_ST}
\end{figure}
%
%%%%%%%%%
%
\Reffig{compare_LHC_DD_SI_ST} illustrates the impact of interference effects for the example of the SI DM-proton cross section, $\sisi$.
For a positive sign of the $g_q^V$ coupling, no cancellation effects occur at LO, and the NLO contributions constitute only a small correction to the LO results. However, when the sign of the $g_q^V$ coupling is taken negative, the interference pattern of \refeq{eq:interference-nlo} has a strong impact on the theoretical predictions. Because the LO contribution is artificially small, the interference of the tree-level contributions with the one-loop corrections as well as the squared one-loop term are the dominant contributions at NLO, and much larger than the pure LO result. Consequently, the NLO corrections strongly modify the LO results.
It can furthermore be seen that the scale dependence is much more pronounced in the case of a suppression of the LO contribution than in the case of no suppression, denoting a strong perturbative uncertainty and a larger relevance of higher-order effects.

Analogous effects are found for the SD DM-neutron cross section $\sisd$, as shown in \reffig{compare_LHC_DD_SD_ST}. We note, however, that due to an opposite sign of the tree-level Wilson coefficient $c^{t,\tree}_A$ compared to $c^{t,\tree}_V$, see \refeq{eq:LOwilt}, the effects are inverted for positive and negative signs of the $g_q^A$ coupling.

%%%%%%%%%%%%%%%%%%%%%%%%%%%%%%%%%%%%%%%%%%%
\section{Summary and conclusions}\label{s:conclusion}
%%%%%%%%%%%%%%%%%%%%%%%%%%%%%%%%%%%%%%%%%%%
In this work we have presented precise predictions for the DM-nucleon cross sections that are searched for in direct detection experiments. For our work we focused on simplified models for the DM-nucleon interaction with particular focus on a $t$-channel and an $s+t$-channel model, the latter including the interaction terms of a simple $s$-channel model. In order to identify predictions obtained within these models with experimental results that are presented in terms of EFT expressions, we performed a matching of simplified model amplitudes to corresponding EFT quantities in terms of Wilson coefficients and effective operators. NLO-QCD corrections for the relevant Wilson coefficients and effective operators were computed explicitly.

We then performed a detailed phenomenological analysis of SI and SD DM-nucleon cross sections in the framework of representative $t$-channel and $s+t$-channel models. After a careful assessment of the theoretical uncertainties associated with the perturbative treatment of the hard scattering we provided a comparison of limits from the LHC and from the CRESST and XENON direct detection experiments. While the impact of NLO corrections on LHC exclusion limits is rather mild in the framework of genuine $t$-channel models, larger effects can be obtained in models featuring the interference of $t$- and $s$-channel contributions.
Whereas LHC limits from mono-jet searches exclude large regions of parameter space for the specific DM models forming the basis of their analyses, limits from direct detection experiments do not rely on any model assumptions. As expected, limits from the CRESST and XENON experiments are somewhat complementary, since CRESST is designed to be particularly sensitive to light DM candidates, while XENON performs best in the domain of larger DM masses.

%%%%%%%%%%%%%%%%%%%%%%%%%%%%%%%%%%%%%%%%%%%
\section*{Acknowledgements}
%%%%%%%%%%%%%%%%%%%%%%%%%%%%%%%%%%%%%%%%%%%
The authors would like to thank Marco Stratmann and Werner Vogelsang for valuable discussions. Part of this work was performed on the high-performance computing resource bwForCluster NEMO with support by the state of Baden-W\"urttemberg through bwHPC and the German Research Foundation (DFG) through grant no INST 39/963-1 FUGG. The Feynman diagrams of this paper have been drawn with \textsc{JaxoDraw 2.0}\cite{Binosi:2003yf,Binosi:2008ig}. All plots have been created with the Python package \textsc{Matplotlib} \cite{Hunter:2007}.

\clearpage
\appendix
%%%%%%%%%%
\section{Explicit NLO results in the effective field theory formalism}\label{s:NLO-expressions}
%%%%%%%%%%
In this appendix, we list all expressions that are required to calculate the $\mathcal{O}(\alphas)$ contributions to the Wilson coefficients for the simplified DM models considered in this work. For clarity, we separate the contributions according to the type of mediator-fermion interaction $j = S, V, A, T$ 
and the topology of the one-loop diagrams by introducing labels for propagator ($\circ$), vertex ($\triangle$), and box corrections ($\square$) as well as counterterm contributions ($\CT$). For symbolic manipulations, we have used the \textsc{Mathematica} packages \textsc{FeynCalc} \cite{Mertig:1990an,Shtabovenko:2016sxi,Shtabovenko:2020gxv} and \textsc{Package-X} \cite{Patel:2015tea}.

In our calculations, we are using conventional dimensional regularization with $d = 4 - 2\varepsilon$ dimensions and anticommuting $\gamma_5$. We use a hybrid on-shell/$\MSbar$ renormalization scheme: While we apply the on-shell scheme for the quark field-strength renormalization, the $\MSbar$ scheme is used for the renormalization of couplings and mediator masses. 

We are working in the non-relativistic limit ($q^2 \to 0$), where the center-of-mass energy squared of the DM-quark system is given as $\hat s = \left(\mchi + m_q\right)^2$ and the square of the momentum transfer between incoming DM particle and outgoing quark reduces to \mbox{$\hat u=\left(\mchi - m_q\right)^2$}.
In the following, we list the expressions for DM-quark scattering and keep the explicit $\hat u$ dependence. Corresponding expressions for anti-DM-quark scattering can easily be obtained thereof by applying the crossing relation $\hat u \leftrightarrow \hat s$.

%%%%%%%%%%
\subsection{Wilson coefficients for the $s$-channel model}\label{s:app:schannel-corrections}
%%%%%%%%%%
In the $s$-channel model, the only tensor structures appearing in the tree-level matrix elements that contribute to DM-quark scattering are the ones corresponding to the vector and axial-vector currents. Imposing the matching condition of \refeq{eq:wil-coeff-tree} leads to the following Wilson coefficients:
\begin{equation}
	\label{eq:LOwils}
	\begin{split}
		c_S^{s,\tree} &= 0\,,\\
		c_V^{s,\tree} &= -\frac{g^V_q g^V_\chi}{\MV^2}\,,\\
		c_A^{s,\tree} &= -\frac{g^A_q g^A_\chi}{\MV^2}\,,\\
		c_T^{s,\tree} &= 0\,.
	\end{split}
\end{equation}
At the one-loop level, there are only contributions originating from corrections to the quark-antiquark-mediator vertex, since the mediator does not carry a color charge. The relevant corrections are of the form
\begin{equation}
	\label{eq:NLOstriangle}
	\begin{split}
		c_{S,\triangle}^{s,\oneloop} &= 0\,,\\
		c_{V,\triangle}^{s,\oneloop} &= 3\left(\Delta + \ln\frac{\mur^2}{m_q^2}\right) + 4\,,\\
		c_{A,\triangle}^{s,\oneloop} &= c_{V,\triangle}^{s,\oneloop} - 2\,,\\
		c_{T,\triangle}^{s,\oneloop} &= 0\,,
	\end{split}
\end{equation}
with $\Delta = \frac{1}{\varepsilon} - \gamma_{\text{E}} + \ln(4\pi)$, where $\gamma_{\text{E}}$ is the Euler--Mascheroni constant $\gamma_{\text{E}} \approx 0.57721$. Note that here and in the following, the $1/\varepsilon$ term generally receives contributions both from UV and IR divergences. It turns out that  the counterterm contributions are the same for both the vector and axial-vector currents, and furthermore they are exactly opposite to the corrections of the vector current:
\begin{equation}
	\label{eq:NLOsCT}
	\begin{split}
		c_{S,\CT}^{s,\oneloop} &= 0\,,\\
		c_{V,\CT}^{s,\oneloop} &= -c_{V,\triangle}^{s,\oneloop}\,,\\
		c_{A,\CT}^{s,\oneloop} &= c_{V,\CT}^{s,\oneloop}\,,\\
		c_{T,\CT}^{s,\oneloop} &= 0\,.
	\end{split}
\end{equation}
We can already see at this point that there is no correction to the vector current.

After summing all the one-loop contributions, we also need to take into account the one-loop operator corrections as dictated by the matching condition of \refeq{eq:wil-coeff-oneloop}. Since the corrections to the vector and axial-vector operators are equivalent to the ones in the $s$-channel model (see below in \ref{s:app:operator-corrections}), they cancel each other, and at $\mathcal{O}(\alphas)$ we find no contributions to the Wilson coefficients of the discussed operators for this model,
\begin{equation}
	c_j^{s,\oneloop} = \frac{\alphas}{4\pi}C_F\left(c_{j,\triangle}^{s,\oneloop} + c_{j,\CT}^{s,\oneloop} - o_j^\oneloop\right) c_j^{s,\tree} = 0\,,
\end{equation}
for all tensor structures $j = S, V, A, T$ and with the color factor $C_F = 4/3$. We note that, as mentioned before, while the sum of the vertex correction and counterterm contribution to the vector current cancels independently of the matching, the corresponding contribution to the axial-vector current is non-zero and vanishes only after including the operator correction.

%%%%%%%%%%
\subsection{Wilson coefficients for the $t$-channel model}
%%%%%%%%%%
Also in the $t$-channel model, only the vector and axial-vector operators contribute at tree level. We have used Fierz identities (see, e.g., Ref.~\cite{Dent:2015zpa}) to arrange the spinor fields in the same order as in the operators of \refeq{eq:leff}. From the tree-level matching condition, we then obtain the following Wilson coefficients:
\begin{equation}
	\label{eq:LOwilt}
	\begin{split}
		c_S^{t,\tree} &= 0\,,\\
		c_V^{t,\tree} &= \frac{\lambda_{Q_L}^2}{8\left(\hat u - \MQ^2\right)} = c^{t,\tree}\,,\\
		c_A^{t,\tree} &= -c^{t,\tree}\,,\\
		c_T^{t,\tree} &= 0\,.
	\end{split}
\end{equation}
Here we have introduced a common tree-level factor, $c^{t,\tree}$, which will also appear in the one-loop expressions for each tensor structure. It is quoted  for our choice of model parameters, i.e.\ for $t$-channel mediators which only couple to left-handed quarks. A transition to the general case is straightforward, i.e.\ by replacing $\lambda_{Q_L}^2$ by a sum of the allowed couplings.

At $\mathcal{O}(\alphas)$, we now have several contributions coming from propagator, vertex, and box corrections. We write these contributions in terms of scalar and tensor integrals in the notation of Ref.~\cite{Patel:2015tea} and use the \textsc{Mathematica} extension \textsc{Package-X} to obtain analytical expressions for the loop integrals. All one-loop integrals emerging in our calculation can be expressed in terms of a small set of master integrals that below we abbreviate as
\begin{align*}
	A_0 &= A_0(\mchi^2)\,,\\
	B_0 &= B_0(\hat u; 0, \MQ^2)\,,\\
	C_i &= C_i(m_q^2, \mchi^2, \hat u; 0, m_q^2, \MQ^2)\,,\\
	D_{i,ij} &= D_{i,ij}(m_q^2, \mchi^2, \mchi^2, m_q^2, \hat u, 0; 0, m_q^2, \MQ^2, m_q^2)\,.
\end{align*}
For the three-point integrals, we encounter $C_i$ with $i=0,1,2$, while in the case of four-point integrals, those of type $D_i$ and $D_{ij}$ with $i,j=0,1,2,3$ emerge.

The contributions from the propagator correction can then be written as:
\begin{equation}
	\begin{split}
		c_{S,\circ}^{t,\oneloop} &= 0\,,\\
		c_{V,\circ}^{t,\oneloop} &= \frac{2\left(\hat u + \MQ^2\right)B_0 - A_0}{\hat u - \MQ^2}\,,\\
		c_{A,\circ}^{t,\oneloop} &= -c_{V,\circ}^{t,\oneloop}\,,\\
		c_{T,\circ}^{t,\oneloop} &= 0\,.
	\end{split}
\end{equation}
The contributions from the vertex corrections are:
\begin{equation}
	\begin{split}
		c_{S,\triangle}^{t,\oneloop} &= 4m_q \mchi C_1\,,\\
		c_{V,\triangle}^{t,\oneloop} &= 2\Big[B_0 + 2\left(\hat u - \mchi^2\right)\left(C_0 + C_1 + C_2\right) + 2\left(m_q^2 C_0 + \mchi^2 C_2\right) \Big]\,,\\
		c_{A,\triangle}^{t,\oneloop} &= -c_{V,\triangle}^{t,\oneloop}\,,\\
		c_{T,\triangle}^{t,\oneloop} &= -c_{S,\triangle}^{t,\oneloop}\,.
	\end{split}
\end{equation}
The contributions from the box correction are:
\begin{equation}
	\begin{split}
		c_{S,\square}^{t,\oneloop} &= 4 m_q \mchi \left(\hat u - \MQ^2\right)\left(D_{12} + D_{22} + D_{23} - D_2\right)\,,\\
		c_{V,\square}^{t,\oneloop} &= 2\left(\hat u - \MQ^2\right) \Big\{ 2D_{00} - \mchi^2 D_{22} - m_q^2\big[D_{11} + D_{22} + D_{33}\\
		&\qquad\qquad\qquad~~~ - 2(D_0 + D_1 + D_2 + D_3 - D_{12} - D_{13} - D_{23})\big]\Big\}\,,\\
		c_{A,\square}^{t,\oneloop} &= -2\left(\hat u - \MQ^2\right) \Big\{ 2D_{00} + \mchi^2 D_{22} + m_q^2\big[D_{11} + D_{22} + D_{33}\\
		&\qquad\qquad\qquad\quad~~ + 2(D_0 + D_1 + D_2 + D_3 + D_{12} + D_{13} + D_{23}) \big] \Big\}\,,\\
		c_{T,\square}^{t,\oneloop} &= 4 m_q \mchi \left(\hat u - \MQ^2\right)\left(D_{12} + D_{22} + D_{23} + D_2\right)\,.
	\end{split}
\end{equation}
We also list the counterterm contributions which we have calculated from the renormalization constants of our model:
\begin{equation}
	\begin{split}
		c_{S,\CT}^{t,\oneloop} &= 0\,,\\
		c_{V,\CT}^{t,\oneloop} &= -\left[\Delta \left(\frac{2\hat u + \MQ^2}{\hat u - \MQ^2} + 4\right) + 3\ln\frac{\mur^2}{m_q^2} + 4\right]\,,\\
		c_{A,\CT}^{t,\oneloop} &= -c_{V,\CT}^{t,\oneloop}\,,\\
		c_{T,\CT}^{t,\oneloop} &= 0\,.
	\end{split}
\end{equation}
Eventually, we sum over all the one-loop contributions and multiply by the common factor mentioned above. Additionally, we need to take into account the one-loop operator corrections as dictated by the matching condition of \refeq{eq:wil-coeff-oneloop} to obtain the $\mathcal{O}(\alphas)$ corrections to the  Wilson coefficients of the $t$-channel model:
\begin{equation}
	\label{eq:NLOtoneloop}
	c_j^{t,\oneloop} = \frac{\alphas}{4\pi}C_F\Big(c_{j,\circ}^{t,\oneloop} + c_{j,\triangle}^{t,\oneloop} + c_{j,\square}^{t,\oneloop} + c_{j,\CT}^{t,\oneloop} - o_j^\oneloop\Big) c^{t,\tree}\,,
\end{equation}
for each tensor structure $j = S, V, A, T$.

%%%%%%%%%%
\subsection{Wilson coefficients for the $s + t$-channel model}\label{s:app:stchannel-corrections}
%%%%%%%%%%
For the $s + t$-channel model, corresponding to the sum of the Lagrangians of \refeq{eq:L-sim-mod-s} and \refeq{eq:L-sim-mod-t}, the tree-level and one-loop Wilson coefficients can be readily obtained by summing the contributions of the $s$- and $t$-channel models quoted above. In particular, this means that at the one-loop level, we must add the vertex and counterterm contributions of \refeq{eq:NLOstriangle} and \refeq{eq:NLOsCT} to the one-loop Wilson coefficient of the $t$-channel model, \refeq{eq:NLOtoneloop}, so that the one-loop operator correction is taken into account only once. We find 
\begin{equation}
	\begin{split}
		c_j^{s + t,\tree} &= c_j^{s,\tree} + c_j^{t,\tree}\,,\\
		c_j^{s + t,\oneloop} &= \frac{\alphas}{4\pi}C_F\left(c_{j,\triangle}^{s,\oneloop} + c_{j,\CT}^{s,\oneloop}\right) c_j^{s,\tree} + c_j^{t,\oneloop}\,,
	\end{split}
\end{equation}
for each tensor structure $j = S, V, A, T$. Since the only relevant one-loop corrections in the $s$-channel model either vanish exactly (for the vector operator) or cancel with the corresponding operator corrections (for the axial-vector operator), we can write the one-loop Wilson coefficient also as:
\begin{equation}
	c_j^{s + t,\oneloop} = \frac{\alphas}{4\pi}C_F\Big(c_{j,\circ}^{t,\oneloop} + c_{j,\triangle}^{t,\oneloop} + c_{j,\square}^{t,\oneloop} + c_{j,\CT}^{t,\oneloop}\Big) c^{t,\tree}\,,
\end{equation}
which corresponds to \refeq{eq:NLOtoneloop} without the operator correction.

%%%%%%%%%%
\subsection{One-loop operator corrections}\label{s:app:operator-corrections}
%%%%%%%%%%
As shown in the Feynman diagram of \reffig{feynman_diags_EFT}~(b), there is only one vertex correction plus a corresponding counterterm contributing to each one-loop operator correction in the EFT. In the non-relativistic limit, these one-loop terms are proportional to the corresponding tree-level operators,
\begin{equation}
	\mathcal{O}_j^\oneloop = \frac{\alphas}{4\pi}C_F\, o_j^\oneloop\mathcal{O}_j^\tree\,,
\end{equation}
with
\begin{equation}
	\label{eq:oneloopopcor}
	o_j^\oneloop = o_{j,\triangle}^\oneloop + o_{j,\CT}^\oneloop
\end{equation}
for each tensor structure $j = S, V, A, T$, parameterizing the full one-loop correction and the color factor $C_F$ explicitly factored out.

In the following, we do not consider the one-loop corrections to the scalar and tensor operators, $o_S^\oneloop$ and $o_T^\oneloop$, as they are irrelevant for our calculation: Since they do not appear at tree level in our models, in the matching formula of \refeq{eq:wil-coeff-oneloop}, they are multiplied by vanishing tree-level Wilson coefficients and therefore do not contribute to the cross sections we consider. We note, however, that in general, the one-loop operator corrections do not necessarily have to vanish.

For the vector and axial-vector operators, the expressions for the operator corrections are identical to the one-loop Wilson coefficients in the $s$-channel model (as discussed above in section~\ref{s:app:schannel-corrections}), because of the same Lorentz structure of the couplings. For the vertex correction to the operators we can therefore write:
\begin{equation}
	\begin{split}
		o_{S,\triangle}^\oneloop &= 0\,,\\
		o_{V,\triangle}^\oneloop &= c_{V,\triangle}^{s,\oneloop}\,,\\
		o_{A,\triangle}^\oneloop &= c_{A,\triangle}^{s,\oneloop}\,,\\
		o_{T,\triangle}^\oneloop &= 0\,,
	\end{split}
\end{equation}
and for the counterterms:
\begin{equation}
	\begin{split}
		o_{S,\CT}^\oneloop &= 0\,,\\
		o_{V,\CT}^\oneloop &= c_{V,\CT}^{s,\oneloop}\,,\\
		o_{A,\CT}^\oneloop &= c_{A,\CT}^{s,\oneloop}\,,\\
		o_{T,\CT}^\oneloop &= 0\,,
	\end{split}
\end{equation}
with $c_{V,i}^{s,\oneloop}$, $c_{A,i}^{s,\oneloop}$ as defined in \refeqs{eq:NLOstriangle} and \eqref{eq:NLOsCT}. Summing the contributions to obtain the full one-loop operator correction as in \refeq{eq:oneloopopcor}, we see that the corrections vanish for the vector operator, and lead to a finite contribution for the axial-vector operator:
\begin{equation}
	\begin{split}
		o_V^\oneloop &= 0\,,\\
		o_A^\oneloop &= -2\,.
	\end{split}
\end{equation}

\end{document}